# How Insight Emerges in a Distributed, Content-addressable Memory

Liane Gabora
and
Apara Ranjan

University of British Columbia

Address for Correspondence:
L. Gabora <liane.gabora@ubc.ca>
Department of Psychology
University of British Columbia
Okanagan campus, 3333 University Way
Kelowna BC, V1V 1V7
CANADA

We begin this chapter with the bold claim that it provides a neuroscientific explanation of the magic of creativity. Creativity presents a formidable challenge for neuroscience. Neuroscience generally involves studying what happens in the brain when someone engages in a task that involves responding to a stimulus, or retrieving information from memory and using it the right way, or at the right time. If the relevant information is not already encoded in memory, the task generally requires that the individual make systematic use of information that is encoded in memory. But creativity is different. It paradoxically involves studying how someone pulls out of their brain something that was never put into it! Moreover, it must be something both new and useful, or appropriate to the task at hand. The ability to pull out of memory something new and appropriate that was never stored there in the first place is what we refer to as the *magic of creativity*. We will see that (like all magic acts) it isn't really magic after all; there is a clever trick behind it.

The difficulty of achieving a neuroscientific account of creativity goes far beyond the problem of getting people to be creative on demand. Even if we are so fortunate as to determine which areas of the brain are active and how these areas interact during creative thought, we will not have an answer to the question of how the brain comes up with solutions and artworks that are new and appropriate. Although standard technologies for investigating brain activity such as fMRI, EEG, and PET may have much to tell us about creativity, in their current state of development they focus at too high a level to explain the magic of creativity.

On the other hand, since the representational capacity of neurons emerges at a level that is higher than that of the individual neurons themselves, i.e. there is no 'grandmother neuron' that always responds to your grandmother, or to Halle Berry, and nothing else, the inner workings of neurons is too low a level to explain the magic of creativity. Although it is a little unsatisfying, even if we do not yet know everything there is to know about the conditions under which individual neurons form new dendrites, or what temporary or permanent intracellular changes take place in response to novel stimuli, in order to 'go for the gold', we assume these things happen *somehow,* and move up a level.

Thus we look to a level that is midway between gross brain regions and neurons. Since creativity generally involves combining concepts from different domains, or seeing old ideas from new perspectives, we focus our efforts on the neural mechanisms underlying the representation of concepts and ideas. Thus we ask questions about the brain at the level that accounts for its representational capacity, i.e. at the level of distributed aggregates of neurons (Blasdel & Salama, 1986; Chandrashekharan, 2009; Churchland & Sejnowski, 1992; Dayan & Abbott, 2001; Eliasmith & Anderson, 2003; Hebb, 1949, 1980; Lin et al., 2005; Lin, Osan, & Tsien, 2006; Smith & Kosslyn, 2007).

We lack methods that permit the necessary spatial resolution to zero in on distributed aggregates of neurons as someone creates, and the necessary temporal resolution to see what these neurons are doing, so in choosing to focus at this level we are in a sense we are groping in the dark. We focus at this level not because the proper tools already exist—i.e., not because that's where the light is—but because that's where we need to look—i.e., that's where we think the quarter is. In the absence of the ability to directly manipulate and experiment with events at the level midway between gross brain regions and neurons during real *in vivo* bouts of creative thinking, computational models, and a little detective work, play roles in our account of the mechanisms underlying creative insight. We believe that we have located something that has the size, shape, and feel of the real quarter. It explains what is commonly referred to as 'jumping out of the box', or what Boden (1990) refers to as *transformational creativity:* creativity that involves not just exploring but changing the space of possibilities. The bulk of this account was put forward over a decade ago (Gabora, 2001) and refined since (Gabora, 2002, 2010, under revision); here we summarize the key elements, and frame them in terms of more recent neuroscientific findings.

## The Ingenious Way that Representations are Encoded in Memory

We said that creativity involves pulling something out of your brain that was never put into it. Nevertheless it is generally assumed that what gets pulled out bears *some* relationship to knowledge and experiences encoded in memory before the creative act took place. But although we spend 15 to 50% of our time engaged in mindwandering, recalling and playing with existing knowledge and ideas in an undirected manner (Smallwood & Schooler, 2006), little of this is what we would call creative; only rarely would one of the thoughts one has during a bout of mindwandering qualify as an insight (Andrews-Hanna, Reilder, Huang, & Buckner, 2010; Signer & Antrobus, 1963). So one question is: what is going on above and beyond the usual tinkering and rearranging that occurs in everyday mindwandering when a genuinely creative idea emerges? A second question is the following. There are a potentially infinite number of different ways of tweaking what we know to come up with something new. How is it that people so often manage to hit on ideas that are *just right?* We believe that the answer to these questions can be obtained by looking at the ingenious way that one's history of experiences is encoded in memory. A brain contains information that was never explicitly stored there but that is *implicitly* present nonetheless. We propose that this implicitly present information enables one to go beyond what one knows without resorting to trial and error.

We take as a starting point some fairly well established characteristics of memory. Human memories are encoded in neurons that are sensitive to ranges (or values) of *microfeatures* (Churchland & Sejnowski, 1992; Churchland, Sejnowski, & Arbib, 1992; Smolensky, 1988). For example, one might respond to lines of a particular orientation, or the quality of honesty, or quite possibly something that does not exactly match an established term (Miikkulainen, 1997). Although each neuron responds maximally to a particular microfeature, it responds to a lesser extent to related microfeatures, an organization referred to as *coarse coding* (Hubel & Wiesel, 1965). Not only does a given neuron participate in the encoding of many memories, but each memory is encoded in many neurons. For example, neuron *A* may respond preferentially to sounds of a certain frequency, while its neighbor *B* responds preferentially to sounds of a slightly different frequency, and so forth. However, although *A* responds *maximally* to sounds of one frequency, it responds to a lesser degree to sounds of a similar frequency. The upshot is that an item in memory is *distributed* across a cell assembly that contains many neurons, and likewise, each neuron participates in the storage of many items (Hebb, 1949; Hinton, McClelland, & Rumelhart, 1986). A given experience activates not just *one* neuron, nor *every* neuron to an equal degree, but activation is spread across members of an assembly. The same neurons get used and re-used in different capacities, a phenomenon referred to as *neural re-entrance* (Edelman, 1987).

The final key attribute of memory is the following. Memory is said to be *content addressable,* meaning that there is a systematic relationship between the content of a representation and the neurons where it gets encoded. This emerges naturally as a consequence of the fact that representations activate neurons that are tuned to respond to particular features, so representations that get encoded in overlapping regions of memory share features. As a result, they can thereafter be evoked by stimuli that are similar or 'resonant' in some (perhaps context-specific) way (Hebb, 1949; Marr, 1969).

This kind of distributed, content-addressable memory architecture is schematically illustrated in Figure 1. Each circle represents a microfeature that is maximally responded to by a particular neuron. Circles that are close together respond to microfeatures that are similar or related. The large, diffuse region of whiteness indicates the region of memory activated by the current thought or experience. Note that even if a brain does not possess a neuron that is maximally tuned to a particular microfeature, the brain is still able to encode stimuli in which that microfeature predominate, because representations are distributed across *many* neurons.

[INSERT FIGURE 1 ABOUT HERE]

The distributed, content-addressable architecture of memory is critically important for creativity. If it were not distributed, there would be no overlap between items that share microfeatures, and thus no means of forging associations between them. If it were not content-addressable, associations would not be meaningful. The upshot of all this is that representations that share features are encoded in overlapping distributions of neurons, and therefore activation can spread from one to another. Thus representations are encoded in memory a way that takes into account how they are related, *even if this relationship has never been consciously noticed*. This is not earth shattering; indeed it seems fairly obvious with respect to the hierarchical structure of knowledge. We may never have explicitly learned that a white goat is a mammal, but we know it is one nonetheless. It is in this sense that we claimed earlier that people implicitly know more than they have ever explicitly learnt. As we will see, this architecture has implications that extend far beyond issues related to the hierarchical structure of knowledge.

It should be pointed out how different this is from a typical computer memory. In a computer memory, each possible input is stored in a unique address. Retrieval is thus a matter of looking at the address in the address register and fetching the item at the specified location. Since there is no *overlap* of representations, there is no means of creatively forging new associations based on newly perceived similarities. The exceptions are computer architectures that are designed to mimic, or are inspired by, the distributed, content-addressable nature of human memory.

## Forging Unusual Associations through Reconstructive Interference of Memories

A fascinating finding to come out of the early connectionist literature is that in a distributed, content addressable memory, not only do representations that share features activate each other, they sometimes interact in a way that is creative. Even a simple neural network is able to abstract a prototype, fill in missing features of a noisy or incomplete pattern, or create a new pattern on the fly that is more appropriate to the situation than anything it has ever been fed as input (McClelland & Rumelhart, 1986). In fact, similar representations can interfere with one another (Feldman & Ballard, 1982; Hopfield, 1982; Hopfield, Feinstein, & Palmer, 1983). Interestingly, the numerous names for this phenomenon—'crosstalk', 'false memories', 'spurious memories', 'ghosts', and 'superposition catastrophe'—are suggestive of a form of thought that, if not outright creative, involves a departure from known reality. Findings from neuroscience are also highly consistent with this phenomenon; indeed as Edelman (2000) puts it, one does not *retrieve* a stored item from memory so much as *reconstruct* it. That is, an item in memory is never re-experienced in exactly the form it was first experienced, but colored, however subtly, by what has been experienced in the meantime, re-assembled spontaneously in a way that relates to the task at hand (one reason eye-witness accounts cannot always be trusted) (Paterson, Kemp, & Forgas, 2009; Loftus, 1980; Schacter, 2001).

The reconstructive nature of memory, while detrimental in some contexts, is beneficial in others; indeed we claim it underlies what was referred to earlier as the magic of creativity. Cognitive psychologists have long struggled with the question of how minds generate ideas that are both new and useful. Almost universally they have concluded that creativity must be a process of search not so different from what happens in a typical computer; it must involve sifting through possibilities, perhaps tweaking or exploring them, until an acceptable one is selected (e.g. Newell, Shaw & Simon 1957; Newell & Simon 1972; Finke, Ward, & Smith, 1992; Simonton, 1999). But by approaching creativity from a neuroscientific perspective, and specifically focusing on our chosen level midway between brains and neurons, we see that the content-addressable, reconstructive nature of memory enables the brain to accomplish creative acts *without* recourse to a 'search and selection' type explanation.

Because information is encoded in a distributed manner across ensembles of neurons interacting by way of synapses, the meaning of a representation is in part derived from the meanings of other representations that excite similar constellations of neurons; that is why it is sometimes referred to as an *associative memory*. Content addressability ensures that the brain naturally brings to mind items that are similar in some perhaps unexpected or indefinable but useful or appealing way to what is being experienced. Recall that if the regions in memory where two distributed representations are encoded overlap then they share one or more microfeatures. They may have been encoded at different times, under different circumstances, and the correlation between them never explicitly noticed. But the fact that their distributions overlap means that *some* context could come along for which this overlap would be relevant, causing one to evoke the other. There are as many routes by which an association between two representations can be forged as there are microfeatures by which they overlap; *i.e.,* there is room for typical as well as atypical connections. Therefore what gets evoked in a given situation is *relevant*, and that happens *for free*—no search is necessary at all—because memory is content-addressable! The 'like attracts like' principle is deeply embedded into our neural architecture.

Moreover, because memory is distributed and subject to crosstalk, if a situation does come along that is relevant to multiple representations, they merge together, a phenomenon that has been termed *reconstructive interference* (Gabora & Saab, 2011). The multiple items may be so similar to each other that you never detect that the recollection is actually a blend of many items. In this case the distributions of neurons they activate overlaps substantially. Or they may differ in mundane ways, as in everyday mindwandering. Alternatively, they may be superficially different but related in a way you never noticed before. In this case the distributions of neurons they activate overlaps only with respect to only a few features that in the present context happen to be relevant or important. Alternatively, the present experience may infuse recall of a previous experience that is relevant or important with respect to only a few key features. For example, the person who invented waterskiing may have been sitting on a beach thinking about snow skiing. The SKIING representation merged with the WATER representation and the idea of WATERSKIING was born. Of course the invention of waterskiing could have happened differently. The person could have been thinking about Jesus walking on water and seen a boat go by slowly with a fish being pulled in on a fishing line. In this case waterskiing was born through the merging of the WALK ON WATER representation with the PULLED BY BOAT representation to give MOVE ACROSS WATER PULLED BY BOAT. Note that WATERSKIING (or MOVE ACROSS WATER PULLED BY BOAT) was not waiting in a dormant, predefined state to be selected, nor was it tweaked or mutated in a trial and error manner. Reconstructive interference of implicitly present information enables one to 'go beyond what one explicitly knows' to solve a problem or express oneself creatively. The greater the extent to which the contributing representations differ, the more likely it is to result in transformative creativity as opposed to mere exploratory creativity.

## Resolving Unusual Associations: States of Potentiality and their Actualization

We saw that reconstructive inteference allows us to generate novelty without having to try out lots of possibilities. However, it has a disadvantage. When representations come together for the first time, it is not always clear how they *go* together; indeed the new idea may barely make sense, as is epitomized by the phrase 'half-baked idea'. For example, having the idea of waterskiing is a far cry from knowing concretely how one would really ski on water. In the beach scenario, the inventor must figure out that the waterskiier is pulled by a boat, and does not need poles. In the Jesus scenario, the inventor must figure out that the waterskiier wears skiis. In either case, the insight initially exists in a state of potentiality; it is not yet clear how it could actualize. So the *effortful* aspect of creativity involves not generating, testing, and selecting, but *actualizing potential* (Gabora, 2005), either by thinking the idea through, or trying it out.

At the moment of insight different possible realizations of the idea have not yet been conceived of. However, they are implicitly present in memory in the sense that each possible realization of the idea activates a different but overlapping constellation of neurons that respond to different sets of microfeatures. For example, one potential realization is that the poles are flattened like paddles. Another potential realization, the most effective one it turns out, involves pulling the skiier from a boat.

## How Distributed Should a Memory be for Creative Insight to Take Place?

How much overlap of microfeatures must there be to result in creative insight? At one extreme it could be not distributed at all, like a typical computer memory. If your mind stored each item in just one location as a computer does, then in order for one experience to remind you of a previous experience, it would have to be *identical* to that previous experience. And since the space of possible experiences is so vast that no two ever *are* exactly identical, this kind of organization would be fairly useless. But at the other extreme, if your memory were *fully distributed*, with each item is stored in every location, the crosstalk would be catastrophe; everything would pretty much remind you of everything.

The problem of crosstalk is solved by *constraining* the distribution region. One way to do this in neural networks is to use a radial basis function, or RBF (Hancock, Smith, & Phillips, 1991; Holden & Niranjan, 1997; Lu, Sundararajan, & Saratchandran, 1997; Willshaw & Dayan, 1990). Each input activates a hypersphere (sphere with more than three dimensions) of locations, such that activation is maximal at the center $k$ of the RBF and tapers off in all directions according to a (usually) Gaussian distribution of width $\sigma$. The result is that one part of the network can be modified without affecting the capacity of other parts to store other patterns. A *spiky activation function* means that $\sigma$ is small. Therefore only those locations closest to $k$ get activated, but they are activated a lot. A *flat activation function* means that $\sigma$ is large. Therefore locations relatively far from $k$ still get activated, but no location gets *very* activated.

In the brain it is the principle of course coding that ensures that distributions are constrained. Because neurons respond most reliably to one particular feature and less reliably to similar features, the region of activation falls midway between two extremes—not distributed at all (a one-to one correspondence between each input and each neuron) and fully distributed (each input activates every neuron). It is because not one neuron, nor every neuron, but a subset of neurons is activated, that one can generate a stream of coherent yet potentially creative thought (Gabora, 2001). The more detail with which items have been encoded in memory, the greater their potential overlap with other items, and the more routes by which one can make sense of the present in terms of the past or engage in creative thinking.

## Insight, Contextual Focus, and Neurds

With flat activation, items are evoked in detail, or multiple items are evoked at once, items with overlapping distributions of microfeatures. Thus flat activation is conducive to forging remote associations amongst items not usually thought to be related, or detecting relationships of correlation. Indeed flat activation would be expected to result in the flat associative hierarchies characteristic of highly creative people (Mednick, 1962). With spiky activation, items are evoked in a compressed form, and few are evoked at once. Thus it is conducive to mental operations on those items, or deducing relationships of causation. Indeed spiky activation would be expected to result in the spiky associative hierarchies characteristic of uncreative people Mednick (1962).

One would imagine that there would be situations where spiky activation would be useful—as when one needs to stay focused, and access remote associations would be distracting, and other situations where flat activation would be useful—as when conventional problem solving methods are not working.

It has long been thought that there are two modes of thought, sometimes referred to as associative and analytic (Freud, 1949; Guilford, 1950; James, 1890) and that we shift between along a continuum between these two extremes depending on the situation we are in (Gabora, 2002; Gabora, 2003). The capacity to shift between the two modes of thought is sometimes referred to as *contextual focus*, because a change from one mode to the other is brought about by the context, through the focusing or defocusing of attention. This is related to *dual process theory*, the idea that cognition employs both implicit and explicit ways of learning and processing information (Chaiken & Trope, 1999; Evans & Frankish, 2009), since analytic thought is believed to involve processing of explicit information whereas associative thought is believed to involve processing of implicit information. Thus contextual focus entails not just the capacity for both associative and analytic thought, but the capacity to adjust the mode of thought to match the demands of the situation. It seems reasonable that we engage in contextual focus using a mechanism akin to varying the size of the RBF: spontaneously tuning the spikiness of the activation function in response to the situation.

What neural mechanisms might underlie the capacity to shift between associative and analytic modes of thought? It has been shown that the cell assembly involved in the encoding of a particular experience is made up of multiple groups of collectively co-spiking neurons referred to as *neural cliques* (Lin et al., 2005; Lin, Osan, & Tsien, 2006). Techniques that enable their patterns of activation to be mathematically described, directly visualized, and dynamically deciphered, reveal that some cliques respond to situation-specific elements of an experience (*e.g.,* where it took place), while others respond to characteristics of varying degrees of generality or abstractness. These range from the type of experience (*e.g.,* being dropped) to characteristics common to many types of experience (*e.g.,* anything dangerous). This has been depicted as a pyramid in which cliques that respond to the most context-specific elements are at the top, and those that respond to the most general elements are at the bottom (Lin et al., 2005).

We can now make a reasonable hypothesis for what is happening at the level of neural cliques during creative thought. Each successive thought activates recruitment of more or fewer neural cliques, depending on the nature of the problem, and how far along one is in solving it. Two well-established phenomena help ensure that a particular thought doesn't recursively reactivate itself. First, if the same neurons are stimulated repeatedly they become refractory. For the duration of this refractory period they cannot fire, or their response is greatly attenuated. Second, they 'team play'; a response is produced by a cooperative group of neurons such that when one is refractory another is active. Since the situation-general neurons and the situation-specific neurons are not responding to the same aspects of the situation, they are not entering and leaving their refractory periods in synchrony, making it unlikely that one will think the same, identical thought over and over again (although over a longer time frame one may repeatedly cycle back to it).

Returning to Figure 1 we can get a schematic picture of how memory is activated by a particular thought. Recall that the degree to which any given region of memory is activated by the current thought or experience is indicated by the degree of whiteness. The white area thus represents the active cell assembly composed of one or more neural cliques, indicated by dashed gray circles. The further a neuron is from the center of the white region, the less activation it not only *receives* from the current instant of experience but in turn *contributes* to the next instant, and the more likely its contribution is cancelled out by that of other simultaneously active locations. Using neural network terminology, we say the broader the region affected, the flatter the activation function, and the narrower the affected region, the spikier the activation function. Figure 1 portrays the state of someone sitting on a beach thinking about snow skiing in an analytic state of mind. The white region is narrow because it is activated in an analytic mode of thought. It includes only neurons that respond to typical features of skiing such as the flatness of the skis and pointiness of the poles.

In a state of defocused attention more aspects of a situation get processed; the set of activated microfeatures is larger, and thus the set of potential associations one could make is larger. Figure 2 shows the state of mind of someone sitting at the beach thinking about snow skiing, but here the activation function is flat. Recruitment of neural cliques that respond to abstract elements of the current thought (e.g. slide across smooth surface) causes the individual to extend the idea of sliding across a surface to the present context of being at the beach. There is reconstructive interference of the skiing concept with this current experience, in which features of skiing (e.g. flatness of skis) are merged with features of water (e.g. that it is liquid not solid). Features of water seem irrelevant to skiing, but they are relevant to inventing a means to ski in the summer.

[INSERT FIGURE 2 ABOUT HERE]

The neural cliques that do not fall within the activated region in Figure 2 but do fall within the activated region in Figure 3 are cliques that *would not* be included in a cell assembly if one were in an everyday relatively convergent mode of thought, but *would* be included if one were in an associative mode of thought. We can refer to them as *neurds*. Neurds respond to microfeatures that are of marginal relevance to the current thought. Neurds do not reside in any particular portion of memory. The subset of neural cliques that count as neurds is defined by context, and shifts constantly. For each situation one might encounter a different group of neurds is involved.

The explanation of insight proposed here follows naturally from the above-mentioned discovery of neural cliques that respond to varying degrees of specificity or generality, and the evidence for contextual focus, as well as the well-established phenomenon that activation of an abstract or general concept causes activation of its instances through spreading activation (Anderson, 1983; Collins & Loftus, 1975).[1] Given that some neural cliques respond to specific aspects of a situation and others respond to more general or abstract aspects, we have a straightforward mechanism by which contextual focus could be executed. In associative thought, with more aspects of a situation taken into account, more neural cliques are activated, including those responding to specific features, those responding to abstract elements, and those *they* activate through spreading activation. Activation flows from the specific instance to the abstract elements it instantiates, to other instances of those abstract elements. The neurds concept thus provides a way of referring to neural cliques that respond to features of these other instances that are not features of the original instance.

It is likely that most of the time, for most individuals, neurds are excluded from activated cell assemblies. Their time to shine comes when one has to break out of a rut. In associative thought, broad activation causes more neural cliques to be recruited, including neurds. This enables the next thought to stray far from the one that preceded it, while retaining a thread of continuity. The associative network can be not just penetrated deeply, but traversed quickly, and there is greater potential for representations to overlap in ways they never have before. Thus the potential to unite previously disparate ideas or concepts is high.

## Example and Analysis of an Instance of Insight

We have examined the relationship between contextual focus and the structure of human memory. This synthesis will now be applied to the analysis of a creative act of a sort that is more artistic than the waterskiing example used previously. In keeping with the view that everyone is creative (Beghetto & Kaufman, 2007; Gardner, 1993; Runco, 2004), the creative act that we analyze is not an earthshaking achievement but a simple event in the life of an everyday person.

The situation that motivates the creative act is the following. Amy, an art student, wants privacy in her bedroom but the only kind of curtains her landlord can afford are ugly and so thick they would block out all the light and her plants would die. She asks around for secondhand curtains trying to solve the situation through a straightforward deductive process. Neural cliques that encode memories of various curtains, and other curtain-like objects, are activated. Neurons that respond to attributes of desirable curtains, such as ‘soft’, ‘hangs (on curtain hooks)’, ‘translucent’, ‘large’, and ‘colorful’, are activated.

Her inability to solve the problem rationally eventually leads to a spontaneous and subconscious defocusing of attention. She enters an associative mode of thought, and her activation function becomes flat, such that the associative structure of her memory is more widely probed. Activation of neurons that respond to attributes such as ‘soft’ that are irrelevant to her goals (of obtaining privacy while retaining sufficient light for the plants) decreases, but activation of neurons that respond to attributes such as ‘translucent’ that are relevant to these remains high. Goal-relevant neurds get recruited moving further down Lin et al.’s (2006) feature-encoding pyramid, and her memory begins to respond to not just specific aspects of her situation (i.e., the need for curtains) but to abstract aspects of her situation (i.e., the need for privacy).

Amy starts considering not just different kinds of curtains, but other attractive ways of covering the window that would allow light to come through. Because memory is content-addressable (i.e., there is a systematic relationship between the content of an item and the locations in memory it activates), neural cliques that respond to 'translucent' and 'colorful', now activated in the context of needing to cover the window, had previously encoded memories of certain acrylic paints that might be opaque enough to provide privacy yet translucent enough to let in light. Activation spreads from neural cliques that respond to 'translucent' and 'colorful' to neural cliques that respond to other aspects of these acrylic paints, causing her to consciously think of them in this new context, resulting in reconstructive interference of CURTAINS and PAINTS to give something like 'CURTAIN-PAINTS: the idea of using paint to accomplish the job of curtains. It has some attributes of curtains (e.g., window coverage) and some attributes of paint (e.g., hardens upon application) as well as the attributes they both share that allowed the association to be made. The association between CURTAINS and PAINTS is new because this particular distributed set of neurons has never been activated together before as an ensemble.

Having hit upon this idea of painting the window, she must determine if it will work in practice. Although in the short run a flat activation function is conducive to creativity, maintaining it would be impractical since the relationship between one thought and the next may be remote; thus a stream of thought lacks continuity. Access to obscure associations might at this point be a distraction. Thus, now she enters a more analytic mode by 'decruiting' the neurds, thereby narrowing the region of memory that gets activated. Thought becomes more logical in character because the activation function becomes spikier, thereby affording finer control; fewer locations release their contents to participate in the formation of the next thought. Experimenting with different paints, colors, and brush styles, Amy finds paints that let in light while obscuring visibility.

Once she knows it will work, the actual painting of the window offers more room for creativity, and she returns to a more associative mode of thought. By shifting back and forth along the spectrum from associative to analytic as needed, the fruits of associative thought become ingredients for analytic thought, and *vice versa*.

## Summary and Conclusions

This chapter provides a tentative but sound and plausible explanation of how the creative process works at the level of distributed ensembles of neurons. We believe that it at this level that one can gain insight into what we call the magic of creativity: the ability to pull out of memory something new and appropriate that was never explicitly stored there in the first place.

The brain is able to do this because of the ingenious way that experiences are laid down in memory. First, representations are *distributed* across assemblies of neurons, sometimes referred to as 'neural cliques'. Each neuron is 'tuned' to respond maximally to a particular microfeature: a particular shade of pink for example, or something more abstract. It responds less reliably to nearby shades of pink. This leads to another basic principle of memory, *coarse coding:* each neuron participates in the encoding of many experiences. Thus a memory of a particular event involves activation of not one neuron but a constellation of them. Memory is also *content-addressable,* meaning there is a relationship between the content of an item, and which neurons respond to it. Thus similar events activate overlapping constellations of neurons. This means that items are encoded in memory not just in terms of their features, but in terms of how they relate to each other; the associative structure of the brain reflects underlying statistical regularities in ones' experiences! It is this implicit knowledge of how things are related, indeed related in ways you may never have consciously noticed, that is called upon in the creative process.

Because of the distributed, content-addressable architecture of memory, multiple items may be evoked simultaneously and merge to give rise to a thought that bears some similarity to these multiple items, but that is identical to none of them. The multiple items may be so similar to each other that you never detect that the recollection is actually a blend of many items. In this case the distributions of neurons they activate overlaps substantially. Or they may differ in mundane ways, as in everyday mindwandering. Alternatively, they may be superficially different but related in a way you never noticed before. In this case the distributions of neurons they activate overlap only with respect to only a few features that in the present context happen to be relevant. This phenomenon has been termed *reconstructive interference,* and the result may be an insight that combines elements of both. The greater the extent to which they differ, the greater the extent to which the insight will appear to be an instance of transformative rather than mere exploratory creativity.

When one is stumped and in need of a creative solution, or in a situation conducive to creative self-expression, one defocuses attention and enters a more associative or divergent form of thought, through a process that is believed to work in a manner similar to flattening the activation function in a neural network. Associative thought causes activation of not just neural cliques that respond to situation-specific aspects of an experience, but neural cliques that respond to general or abstract aspects. Those neural cliques that respond to atypical aspects of the situation, and thus that are activated in associative but not analytic thought, are referred to as *neurds*. Items encoded previously to neurds are superficially different from the present situation yet share aspects of its deep structure. Therefore, the recruitment of neurds increases the probability of forging associations that are seemingly irrelevant yet vital to the creative task. By responding to abstract or atypical features of the situation, neurds effectively draw remote associates into the conceptualization of the task. If an insightful association is made, one may enter a more analytic or convergent mode of thought through a process akin to increasing the spikiness of the activation function, by de-activating neurds. Analytic thought discourages potentially disruptive associations and is thus conducive to simply getting a job done.

It is interesting to consider the long-term consequences of the proclivity to shift readily into a state of defocused attention. More features of any given experience evoke 'ingredients' from memory for the next experience. Some aspects of the external world get ignored because one is busy processing previous material, but if something does manage to attract attention, it tends to be considered from multiple angles before settling into a particular interpretation of it. The end result is that one's understanding of the world becomes increasingly unique, and this uniqueness may be reflected in one's creative output.

The explanation of creativity put forward here is incomplete, particularly with respect to the role of motivation and emotions. Moreover, a complete neuroscientific account of creativity will include an explanation of how events at the level of distributed cell assemblies dovetail with, on the one hand, with activity in particular areas of the brain, and on the other hand, the intracellular events that make the formation of new representations possible. We believe, however, that some of the more mysterious aspects of creativity have been solved.

## Acknowledgments

This research was funded by grants from the Social Sciences and Humanities Research Council of Canada (SSHRC) and the Fund for Scientific Research, Flemish Government of Belgium.

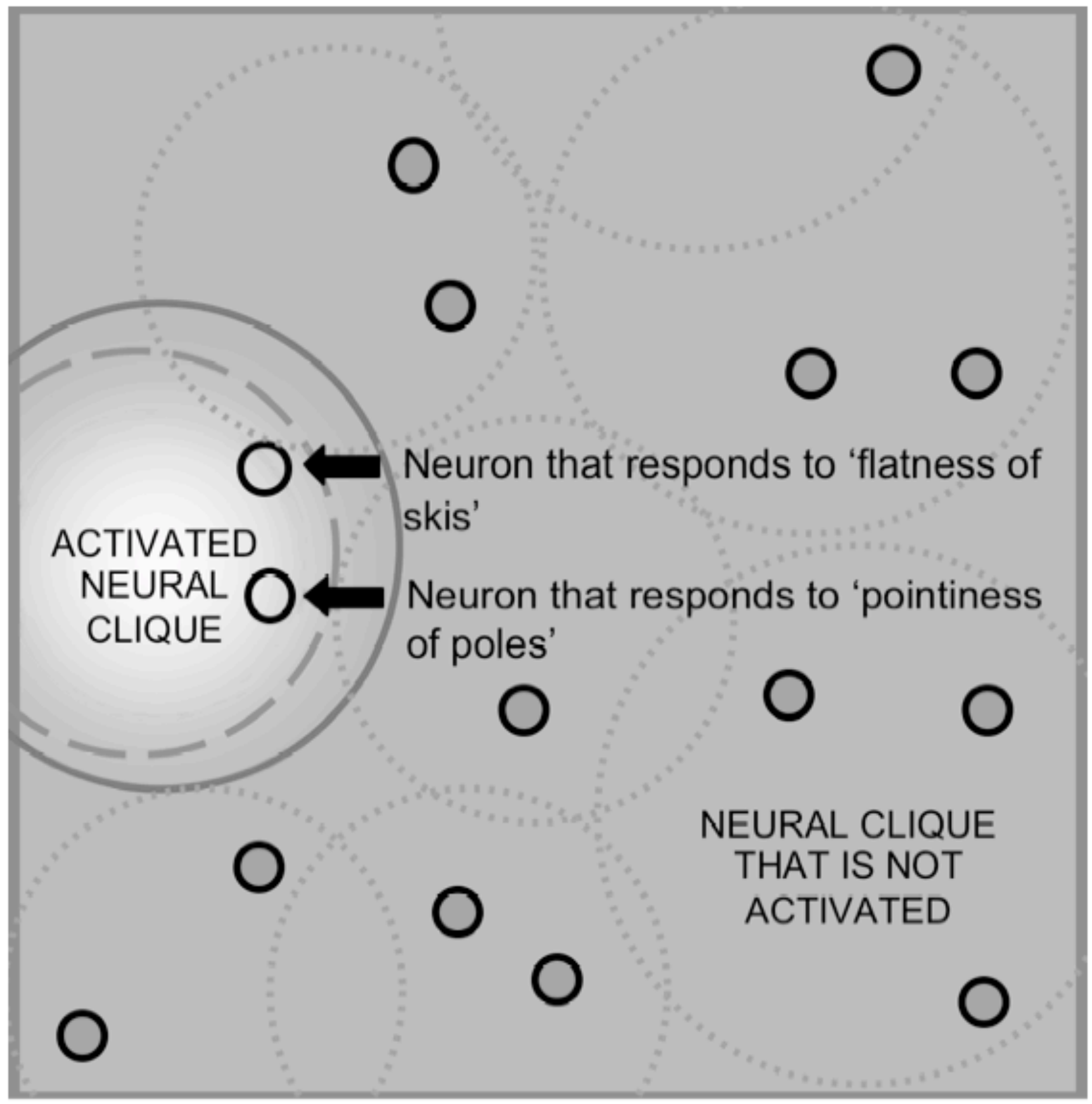

Figure 1. Schematized drawing of a portion of a distributed associative memory activated by the thought of snow skiing in an analytic mode of thought. Each small black-ringed circle represents a feature that a particular neuron responds to. The white region indicates the portion of memory that is activated. The activated cell assembly, indicated by the large grey circle, consists of only one neural clique, indicated by the dashed circle. It is composed of neurons that respond to typical features of snow skiing such as the flatness of the skis and the pointiness of the poles. Non-activated neural cliques are indicated by dotted gray circles.

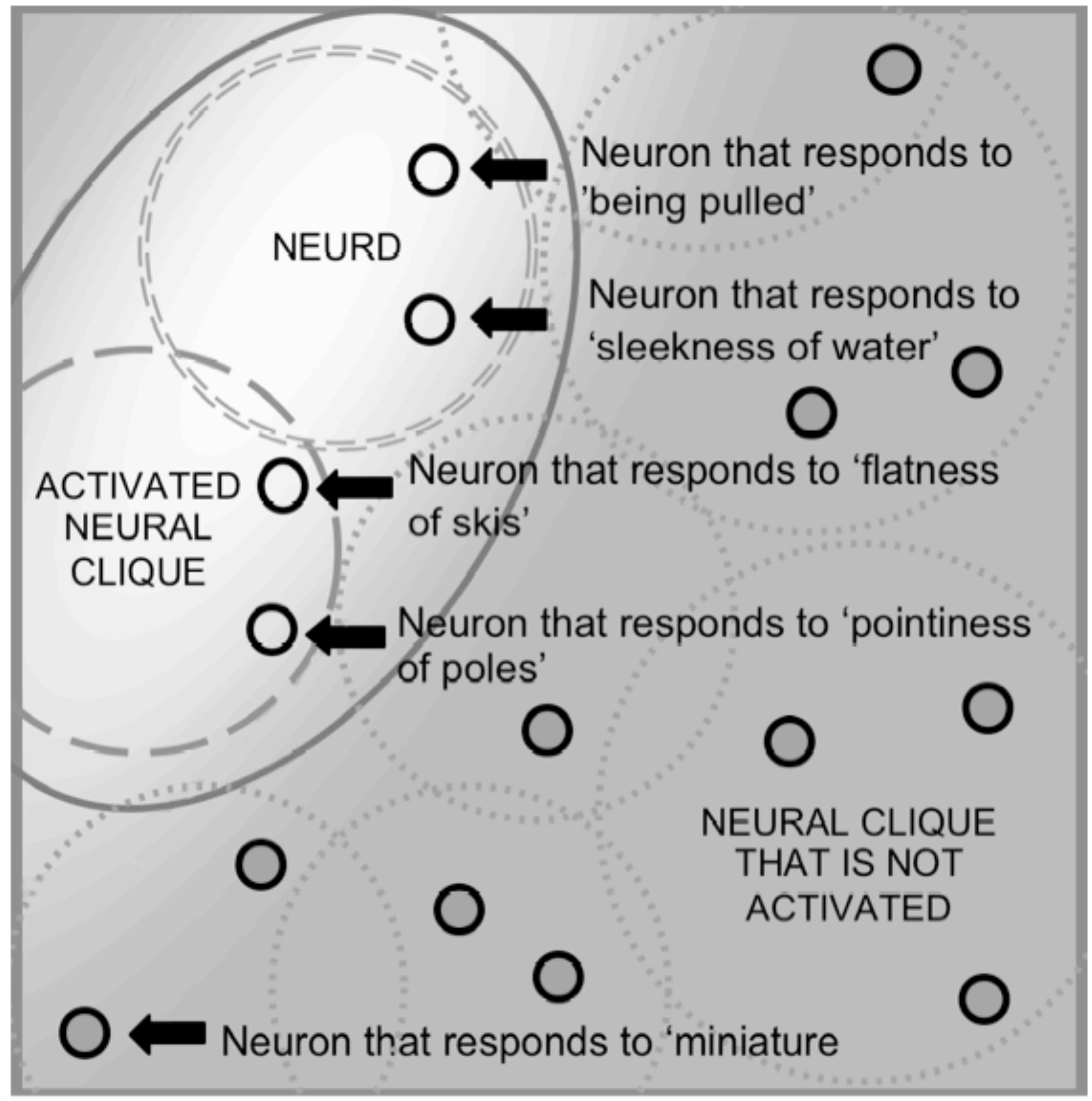

Figure 2. In an associative mode of thought, the portion of memory activated by the thought of snow skiing is larger than it was in an analytic mode, as indicated by the size and diffuseness of the white region. The activated cell assembly, indicated by the large oval, now contains more than one neural clique. The initially activated neural clique is indicated by the dashed circle, and the neurd is indicated by the double circle of dashes. The neurd is composed of neurons that respond to features that are not typical of snow skiing such as 'sleekness of water' but that are relevant to the invention of waterskiing. Note that under a different context, such as the task of making skis for a doll, the neurd might have been a different neural clique, containing the neuron that responds to 'miniature'.

**Notes**

---

[1] Thus for example, based on a set of free association norms data collected from 6,000 participants using over 5,000 words, the probability that, given the word PLANET, the first word that comes to mind is EARTH is .61, and the probability that it is MARS is .10 (Nelson, McEvoy, & Schreiber, 2004). Note that there is some empirical support for an alternative to spreading activation as an explanation for this kind of association data, referred to as 'spooky activation at a distance' (Nelson, McEvoy & Pointer, 2003).